\documentclass[a4paper,fleqn,usenatbib,useAMS]{mnras}

\usepackage{graphicx}	
\usepackage{amsmath}	
\usepackage{amssymb}	
\usepackage{multicol}        
\usepackage{bm}		
\usepackage{pdflscape}	
\usepackage{textgreek}  
\usepackage{float} 
\usepackage[normalem]{ulem} 

\newcommand{\kms}{\,km\,s$^{-1}$} 
\newcommand{\teff}{T$_{\rm eff}$} 
\newcommand{\logg}{\ensuremath{\log g}}

\usepackage[T1]{fontenc}
\usepackage{ae,aecompl}

\usepackage{newtxtext,newtxmath}

\title[Periodic Eclipses of PDS~110]{Periodic Eclipses of the Young Star PDS~110 Discovered with WASP and KELT Photometry}
\author[H. P. Osborn et al]{H. P. Osborn$^{1}$,\thanks{Contact e-mail: \href{mailto:h.p.osborn@warwick.ac.uk}{h.p.osborn@warwick.ac.uk}}
J. E. Rodriguez$^{2}$,
M. A. Kenworthy$^{3}$,
G. M. Kennedy$^{4}$,
E. E. Mamajek$^{5,6}$,
\newauthor
C. E. Robinson$^{7}$,
C. C. Espaillat$^{7}$,
D. J. Armstrong$^{1,8}$,
B. J. Shappee$^{9,10}$,
A. Bieryla$^{2}$,
\newauthor
D. W. Latham$^{2}$,
D. R. Anderson$^{11}$,
T. G. Beatty$^{12,13}$,
P. Berlind$^{2}$,
M. L. Calkins$^{2}$,
\newauthor
G. A. Esquerdo$^{2}$,
B. S. Gaudi$^{14}$,
C. Hellier$^{11}$,
T. W.-S. Holoien$^{12,15,16}$, 
D. James$^{17}$, 
\newauthor
C. S. Kochanek$^{12,15}$,
R. B. Kuhn$^{18}$,
M. B. Lund$^{19}$,
J. Pepper$^{20}$,
D. L. Pollacco$^{1}$,
J. L. Prieto$^{21,22}$,
\newauthor
R. J. Siverd$^{25}$,
K. G. Stassun$^{19,26}$,
D. J. Stevens$^{14}$,
K. Z. Stanek$^{14,15}$,
R. G. West$^{1}$
\\
$^{1}$Department of Physics, University of Warwick, Gibbet Hill Road, Coventry, CV4 7AL, UK\\
$^{2}$Harvard-Smithsonian Center for Astrophysics, 60 Garden St, Cambridge, MA 02138, USA\\
$^{3}$Leiden Observatory, Leiden University, P.O. Box 9513, 2300 RA Leiden, The Netherlands\\
$^{4}$Institute of Astronomy, University of Cambridge, Madingley Road, Cambridge CB3 0HA, UK\\
$^{5}$Jet Propulsion Laboratory, California Institute of Technology, M/S 321-100, 4800 Oak Grove Dr., Pasadena, CA 91109, USA\\
$^{6}$Department of Physics and Astronomy, University of Rochester, Rochester, NY 14627, USA\\
$^{7}$Department of Astronomy, Boston University, One Silber Way, Boston, MA 02215, USA\\
$^{8}$ARC, School of Mathematics \& Physics, Queen's University Belfast, University Road, Belfast BT7 1NN, UK\\
$^{9}$Hubble, Carnegie-Princeton Fellow\\
$^{10}$Las Cumbres Observatory Global Telescope Network, 6740 Cortona Dr., Suite 102, Santa Barbara, CA 93117, USA\\
$^{11}$Astrophysics Group, Keele University, Staffordshire, ST5 5BG, UK\\
$^{12}$Department of Astronomy \& Astrophysics, The Pennsylvania State University, 525 Davey Lab, University Park, PA 16802\\
$^{13}$Center for Exoplanets and Habitable Worlds, The Pennsylvania State University, 525 Davey Lab, University Park, PA 16802\\
$^{14}$Department of Astronomy, The Ohio State University, Columbus, OH 43210, USA\\
$^{15}$Center for Cosmology and AstroParticle Physics (CCAPP), The Ohio State University, 191 W.Woodruff Ave., Columbus, OH 43210, USA\\
$^{16}$US Department of Energy Computational Science Graduate Fellow\\
$^{17}$Astronomy Department, University of Washington, Box 351580, Seattle, WA 98195, USA\\
$^{18}$South African Astronomical Observatory, PO Box 9, Observatory 7935, South Africa\\
$^{19}$Department of Physics and Astronomy, Vanderbilt University, 6301 Stevenson Center, Nashville, TN 37235, USA\\
$^{20}$Department of Physics, Lehigh University, 16 Memorial Drive East, Bethlehem, PA 18015, USA\\
$^{21}$Nucleo de Astronoma de la Facultad de Ingeniera, Universidad Diego Portales, Av. Ejercito 441, Santiago, Chile\\
$^{22}$Millennium Institute of Astrophysics, Santiago, Chile\\
$^{23}$Carnegie Observatories, 813 Santa Barbara Street, Pasadena, CA 91101, USA\\
$^{24}$Hubble, Carnegie-Princeton Fellow\\
$^{25}$Las Cumbres Observatory Global Telescope Network, 6740 Cortona Dr., Suite 102, Santa Barbara, CA 93117, USA\\
$^{26}$Department of Physics, Fisk University, 1000 17th Avenue North, Nashville, TN 37208, USA\\
}
\date{Last updated 2017 Jan 31; in original form 2017 January 31}

\pubyear{2017}

\begin{document}
\label{firstpage}
\maketitle

\begin{abstract}
We report the discovery of eclipses by circumstellar disc material associated with the young star PDS 110 in the Ori OB1a association using the SuperWASP and KELT surveys. 
PDS 110 (HD 290380, IRAS 05209-0107) is a rare Fe/Ge-type star, a $\sim$10 Myr-old accreting intermediate-mass star showing strong infrared excess (L$_{\rm IR}$/L$_{\rm bol}$ $\simeq$ 0.25). 
Two extremely similar eclipses with a depth of $~$30\% and duration $\sim$25 days were observed in November 2008 and January 2011.
We interpret the eclipses as caused by the same structure with an orbital period of $808\pm2$ days.
Shearing over a single orbit rules out diffuse dust clumps as the cause, favouring the hypothesis of a companion at \textasciitilde2AU.
The characteristics of the eclipses are consistent with transits by an unseen low-mass ($1.8-70M_{Jup}$) planet or brown dwarf with a circum-secondary disc of diameter $\sim$0.3 AU.
The next eclipse event is predicted to take place in September 2017 and could be monitored by amateur and professional observatories across the world.
\end{abstract}




\section{Introduction}

The revolution in high-resolution imaging at both near infrared (e.g. SPHERE, \cite{Beuzit:2008}; GPI \cite{Macintosh:2014}) and sub-millimeter wavelengths (ALMA, \citet{Wootten:2009}) is providing new insights into the era of planet formation \citep{Thalmann:2015, Rapson:2015, ALMA:2015}.
This includes structure in circumstellar discs such as rings, spirals and gaps \citep[e.g.][]{Pinte:2015,Stolker:2016}.

The inner regions of discs (\textasciitilde AU) are still too small to be directly imaged.
The transit of dusty circumstellar material in front of a star, however, allows us to resolve the structure of eclipsing material at a resolution limited only by the stellar diameter ($\sim$0.005-0.02\,au).
Eclipses have been previously used to detect inner ring edges in circum-secondary discs ($\epsilon$ Aur \cite{Carroll:1991}; EE Cep, \cite{Galan:2012}), gas accretion streams from the circumstellar disc \citep[e.g.][]{Bouvier:1999,Cody:2014,Ansdell:2016}, sharp outer disc edges in circumsecondary discs \citep[e.g. KH\,15D;][]{Herbst:2002}, and even ring gaps in putative circumplanetary discs \citep[e.g J1407;][]{Mamajek:2012, Kenworthy:2015}.

Unfortunately, these events are rare, with only a dozen or so such eclipsing objects currently known. 
However, projects like the Wide Angle Survey for Planets \citep[WASP;][]{Pollacco:2006} and the Kilodegree Extremely Little Telescope  \citep[KELT;][]{Pepper:2007, Pepper:2012} provide long baseline, high-precision time series photometry for a large portion of the entire sky. 
The combination of baseline, cadence, precision, and sky coverage make these surveys well-suited to search for these ``Disc Eclipsing'' systems. 
The Disc Eclipse Search with KELT (DESK) survey has been conducting an archival search for these unique systems in the $\sim$4 million KELT light curves \citep{Rodriguez:DESK} and has already led to the discovery and analysis of 6 previously unknown large dimming events including the periodic dimming events around V409 Tau \citep{Rodriguez:2015}, DM Ori \citep{Rodriguez:2016C}, and a $\sim$69 year period analogue to $\epsilon$ Aur, TYC 2505-672-1 \citep{Rodriguez:2016B}.
The OGLE survey of the galactic bulge \citep{Udalski:2003} has also discovered young eclipsing candidates that require follow up \citep[e.g.][]{Scott:2014}.

In this paper, we present the light curve of PDS 110, a young star in the Ori OB1 association, which shows two extended, deep dimming events over durations of
$\sim$25 days, separated by about 808 days. 
We interpret these eclipses as due to the transit of a circumsecondary matter associated with an unseen companion PDS 110b, in a bound Keplerian orbit about PDS 110.
In Section 2 we summarise the properties of PDS 110. 
In Section 3 we present photometric and spectroscopic data obtained for PDS~110. 
%
%
In Section 4 we interpret the spectral and photometric data with main-sequence fitting, a simple SED model and a Gaussian eclipse model of the eclipses.
In Section 5 we discuss the likely mechanism behind the eclipses, and 
in Section 6 we cover the prospects for future observations.

\begin{table*}
\centering
\caption{Stellar Parameters for PDS 110}
\label{tbl:Host_Lit_Props}
\begin{tabular}{llccc}
\hline
  Parameter & Description & Value & Source & Reference(s)\\
			&			  &		  &		   &			\\
$\alpha_{J2000}$ &Right Ascension (RA)& 05:23:31.008			& Tycho-2	& \citet{Hog:2000}	\\
$\delta_{J2000}$ &Declination (Dec)& -01:04:23.68		& Tycho-2	& \citet{Hog:2000}	\\
SpT              & Spectral Type   & keF6\,IVeb & ... & \citet{Miroshnichenko:1999}\\
$U$   & Johnson $U$ & 11.02 & PDS & \citet{Gregorio-Hetem:2002}\\
$B$   & Johnson $B$ & 10.934 $\pm$ 0.005 & ...     & \citet{Miroshnichenko:1999},\citet{Pojmanski:2002}\\
$B_T$ & Tycho B$_T$ magnitude& 11.093 $\pm$ 0.058 & Tycho-2 & \citet{Hog:2000}	\\
$V$   & Johnson $V$ & 10.422 $\pm$ 0.002 & ASAS    & \citet{Pojmanski:2002}\\
$V_T$ & Tycho V$_T$ magnitude& 10.476 $\pm$ 0.048 & Tycho-2 & \citet{Hog:2000}	\\
$g'$ & Sloan g'	 & 10.693 $\pm$ 0.032 & APASS 	& \citet{Henden:2015}\\
$R$  & Cousins $R$ & 10.10 & PDS & \citet{Gregorio-Hetem:2002}\\
$r'$ & Sloan r'	 & 10.285 $\pm$ 0.01  & APASS 	& \citet{Henden:2015}\\
$I$  & Cousins $I$ & 9.77 & PDS & \citet{Gregorio-Hetem:2002}\\
$i'$ & Sloan i'	 & 10.174 $\pm$ 0.017 & APASS 	& \citet{Henden:2015}\\
$J$  & 2MASS magnitude & 9.147 $\pm$ 0.021 & 2MASS & \citet{Cutri:2003}\\
$H$	 & 2MASS magnitude & 8.466 $\pm$ 0.042 & 2MASS & \citet{Cutri:2003}\\
$K_s$ & 2MASS magnitude & 7.856 $\pm$ 0.021 & 2MASS & \citet{Cutri:2003}\\
\textit{WISE1} &WISE 3.4\,$\mu$m band mag & 6.941 $\pm$ 0.035 & WISE &\citet{Cutri:2012}\\
\textit{WISE2} &WISE 4.6\,$\mu$m band mag & 6.474 $\pm$ 0.019 & WISE & \citet{Cutri:2012}\\
\textit{WISE3} &WISE 12\,$\mu$m band mag  & 4.512 $\pm$ 0.016 & WISE & \citet{Cutri:2012}\\
\textit{WISE4} &WISE 22\,$\mu$m band mag  & 1.809 $\pm$ 0.021 & WISE & \citet{Cutri:2012}\\
IRAS 12$\mu$m  &IRAS Flux Density (Jy)&0.558 $\pm$ 0.056 &IRAS& \citet{Helou:1988}	\\
IRAS 25$\mu$m  &IRAS Flux Density (Jy)&1.68 $\pm$ 0.10	  &IRAS&\citet{Helou:1988}\\
IRAS 60$\mu$m  &IRAS Flux Density (Jy)&2.13 $\pm$ 0.15	  &IRAS&\citet{Helou:1988}\\
IRAS 100$\mu$m &IRAS Flux Density (Jy)&	{1.68} & IRAS & \citet{Helou:1988}\\
$\mu_{\alpha}$ & Proper Motion in RA (mas yr$^{-1}$) & 1.146\,$\pm$1.067 & Gaia & \citet{Gaia:2016}\\
$\mu_{\delta}$ & Proper Motion in DEC (mas yr$^{-1}$) & -0.338\,$\pm$\,1.076 & Gaia & \citet{Gaia:2016}\\
Distance & pc & 335\,$\pm$\,13 & Hipparcos & \citet{Hernandez:2005}\\
\hline
\end{tabular}
%
%
\end{table*}

\section{Background on the Star PDS~110 \label{sec:PDS110}}
PDS~110 (also known as HD 290380, IRAS 05209-0107, GLMP 91, 2MASS J05233100-0104237 and TYC 4753-1534-1) has been observed in many photometric \citep{Garcia-Lario:1990,Alfonso-Garzon:2012,Hernandez:2005} and spectroscopic \citep{MacConnell:1982,Torres:1995,Miroshnichenko:1999,Gregorio-Hetem:2002,Rojas:2008} studies.

It was also found to have a signficant infrared excess \citep{Garcia-Lario:1990}, representing roughly 25\% of the total luminosity \citep{Rojas:2008}, which likely has comparable contributions from a disc and a more spherical envelope \citep{Gregorio-Hetem:2002}.
Table~1 summarizes the photometry we will use here. Spectroscopically, it shows H$\alpha$ in emission with an equivalent width of roughly 6\AA\ and LiI~$670.7$~nm in absorption with an equivalent width of 0.08m\AA\ \citep{Gregorio-Hetem:2002}. 
A range of spectral types have been assigned to it (F0 \citet{Cannon:1949}, keF6IVeb \citet{Miroshnichenko:1999}, F7e \citet{Miroshnichenko:1999}).
Rojas et al. (2008) made estimates of the luminosity, mass and age, but used a distance of 600~pc which is significantly larger than its measured distance \citep{Gaia:2016}, leading to overestimates of the mass and luminosity and underestimates of the age.
The foreground extinction is small, with $E(B-V)=0.05$~mag \citep{Miroshnichenko:1999}.

PDS~110 has a GAIA parallax of $2.91\pm0.34$~mas, corresponding to a distance of $345 \pm 40$~pc, and a negligible proper motion ($1.15\pm 1.07$, $-0.34\pm 1.08$)~mas/year \citep{Gaia:2016}.
This distance makes PDS~110 consistent with being a member of the Ori OB1a association which has an estimated distance of $335\pm13$~pc and similarly small mean proper motion of ($0.75\pm0.29$,$-0.18\pm0.22$) \cite{Wu:09}. 
The Ori OB1 association has an estimated age of $7$-$10$~Myr \citep{Briceno:2007, vanEyken:12, Ingleby:2014, Ciardi:14}.
The group contains numerous B stars, but not earlier than B1 \citep{Brown:1994a} suggesting that the age may be slightly higher (10-15~Myr).

\section{Data}
In this section we briefly introduce the photometric and spectroscopic data obtained for PDS-110.
Figure \ref{figure:FullLC} shows full and phase-folded light curves along with views of eclipses observed in 2008 (observed by WASP-North, WASP-S and ASAS) and 2011 (observed by KELT).
If the 2.2 year separation of the eclipses is a period, all other predicted eclipses lie in an observing gap. 
Figures \ref{figure:whtspec} and \ref{figure:SED} show optical and IR spectroscopy and best-fit models.

\subsection{WASP}

The Wide Angle Search for Planets (WASP) is comprised of two outposts, located at the Roque de los Muchachos Observatory on La Palma (WASP-North) and the South African Astronomical Observatory (WASP-South). Each observatory consists of 8 cameras using 200mm f/1.8 lenses and cameras with $2048\times2048$ pixel CCDs, $7.8 \times 7.8$ square degree fields of view, and pixel scales of $13.7\arcsec$ \citep{Pollacco:2006}. Light curves were detrended using a version of the SysRem algorithm developed specifically for WASP \citep{Collier:2006,Tamuz:2005}. 

PDS~110 was observed by both WASP-North and WASP-South with exposure times of 30s and a cadence of 8-10 minutes. In total 49558 observations were taken between UT 2008 January 25 and 2013 February 23. 

Light curves were further cleaned, initially by removing 3-sigma anomalies and regions with high hourly scatter (e.g. with hourly RMS scatter above 3\%). To remove trends present in all nearby stars but not removed by SysRem detrending, nightly linear trends were fitted to the median-divided fluxes of 100 bright and non-varying stars within a 25~arcminute aperture. The target light curve was then divided by these residual trends, improving the average flux rms from 6\% to 3\%.

\subsection{KELT}

The Kilodegree Extremely Little Telescope (KELT) is an all sky, photometric survey of bright stars ($8 < V < 11$) designed to detect transiting planets around bright stars \citep{Pepper:2007, Pepper:2012}. The project is comprised of two telescopes, KELT-North in Sonoita, AZ, USA and KELT-South in Sutherland, South Africa. Both telescopes have a 42 mm aperture, a broad R-band filter, and observed with a 10-20 minute cadence. Using a Mamiya 645-series wide-angle lens with an 80mm focal length (f/1.9), the telescopes have a $26^{\circ}$ $\times$ $26^{\circ}$ field-of-view, and a 23$\arcsec$ pixel scale.

PDS~110 is located in KELT-South field 05 ($\alpha$ =  06hr 07m 48.0s, $\delta$ = $+3^{\circ}$ 00$\arcmin$ 00$\arcsec$). The KELT-South telescope observed PDS~110 from UT 2010 February 28 to UT 2015 April 09, obtaining 2892 observations. For a detailed description of the KELT data acquisition and reduction process, see \citet{Siverd:2012, Kuhn:2016}. The typical per point error is \textasciitilde0.02 \%. 

\begin{figure*}
\centering
\includegraphics[width=\linewidth]{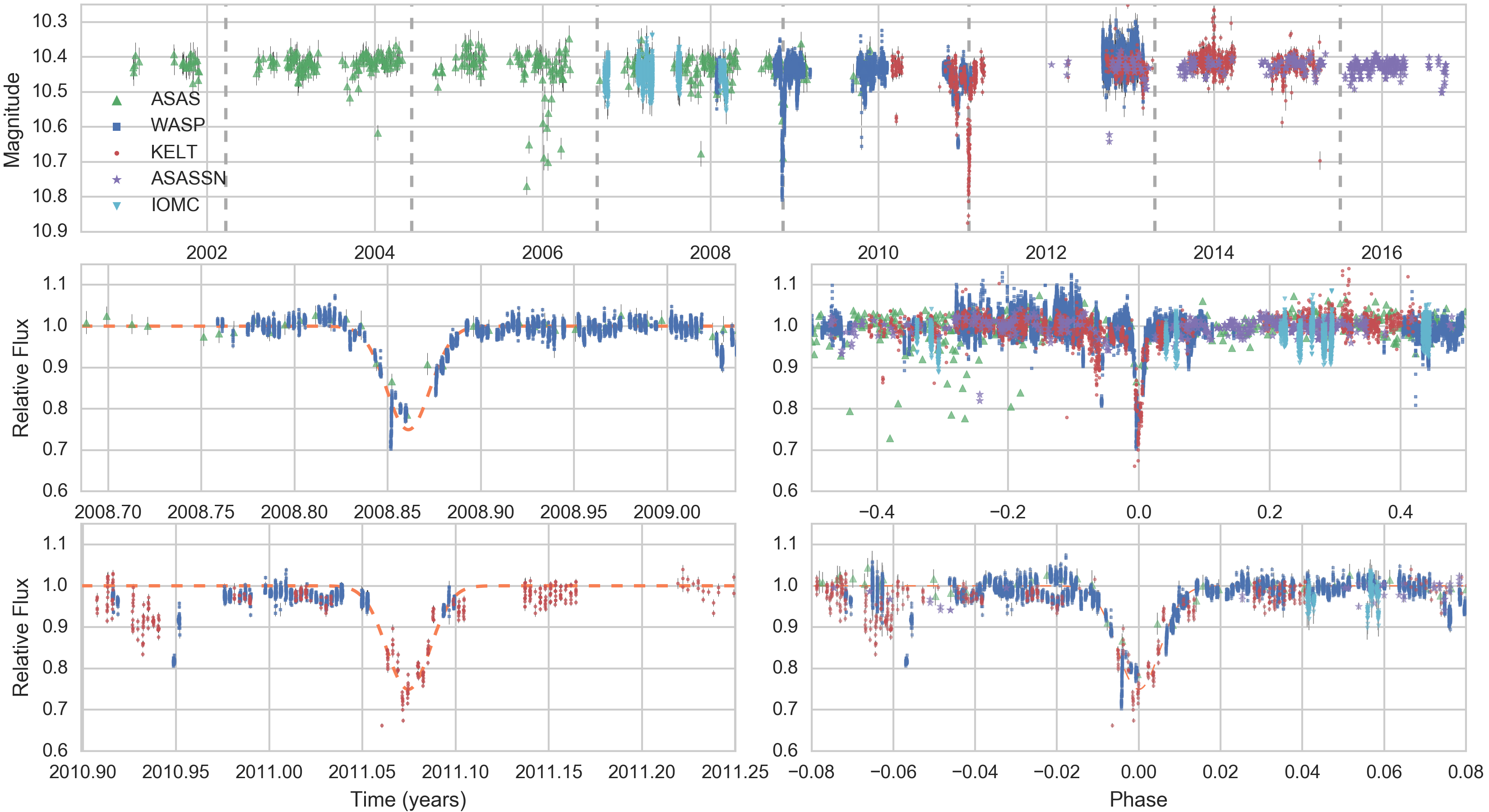}%
\caption{WASP (blue squares), KELT (red circles), ASAS (green triangles), ASAS-SN (purple stars) and IORC (turquoise triagles) observations. Upper figure: plotted from 2002 to 2016. Eclipse times are shown with dashed vertical lines. Lower left figures: individual eclipses in 2008 (upper) and 2011 (lower). Lower right figures: phase-folded light curve with full phase coverage (upper) and zoomed to the eclipse (lower). The best-fit eclipse model (see section 4.3) is overplotted in orange in these cases. In all cases we have applied a vertical offset to the KELT and WASP data to match the quiescent magnitude seen in the ASAS $V$-band data. However, there has been no attempt to place all the data on the same absolute scale.}
\label{figure:FullLC}
\end{figure*}

\subsection{All-Sky Automated Survey (ASAS)}

With the goal of finding and cataloging bright variable stars, the All-Sky Automated Survey (ASAS) obtained photometric observations of a large fraction of the sky. The survey observed simultaneously in two bandpasses ($V$ and $I$) from two observing sites, Las Campanas, Chile and Haleakala, Maui. A detailed description of the survey set up, data acquisition, and reduction pipeline is presented in \citet{Pojamanski:1997}. At each location are two wide-field Minolta 200/2.8 APO-G telephoto lenses and a 2K$\times$2K Apogee CCD. The telescope and camera set up corresponds to a 8.8$^{\circ}\times8.8^{\circ}$ field-of-view and a pixel scale of 13.75$\arcsec$.
PDS~110 was observed from 2001 until 2010\footnote{ASAS data from http://www.astrouw.edu.pl/asas/?page=aasc}. There are 488 ASAS epochs with a typical per-point flux error of 3\%. 

Both KELT and WASP have non-conventional passbands and potential zero point magnitude offsets.
Therefore, to compare them with photometry from other surveys, the out-of-eclipse photometric median was normalised to the out-of-eclipse median of ASAS (Johnson V-band).

\subsection{All-Sky Automated Survey for SuperNovae (ASAS-SN)}
The All-Sky Automated Survey for SuperNovae (ASAS-SN) is monitoring the entire sky every two days down to a $V$-band magnitude of 17. The survey has two separate observing sites, each with four 14 cm Nikon telephoto lenses and 2k $\times$ 2k thinned CCD. The FOV is $4.5\times4.5$ degrees and the pixel scale is 7$\farcs$8. PDS~110 was observed 559 times from 2012 until 2016 with an average per-point rms of 1\%. For a complete description of the observing strategy and reduction process, see \citet{Shappee:2014}.

\subsection{INTEGRAL-OMC}
The INTErnational Gamma-Ray Astrophysics Laboratory (INTEGRAL) \citep{Winkler:2003} is an ESA satellite in orbit since 2002. 
As well as performing gamma ray and X-ray observations, INTERGRAL possesses an Optical Monitoring Camera (OMC, \cite{Mas:2003}), a V-band (500-600nm) imager designed to measure the target's optical brightness and position.
It observed PDS-110 on 14 occasions from 2006 to 2008, taking over 2000 individual flux measurements with an average cadence of 2.7 minutes and a median precision of 1.4\% \citep{Alfonso-Garzon:2012}\footnote{IOMC data accessed from Vizier at http://vizier.cfa.harvard.edu/viz-bin/VizieR?-source=J/A+A/548/A79}.

\subsection{ISIS spectrum}
\begin{figure*}
\centering
\includegraphics[width=\textwidth,clip, trim = 1in 0 1in 0]{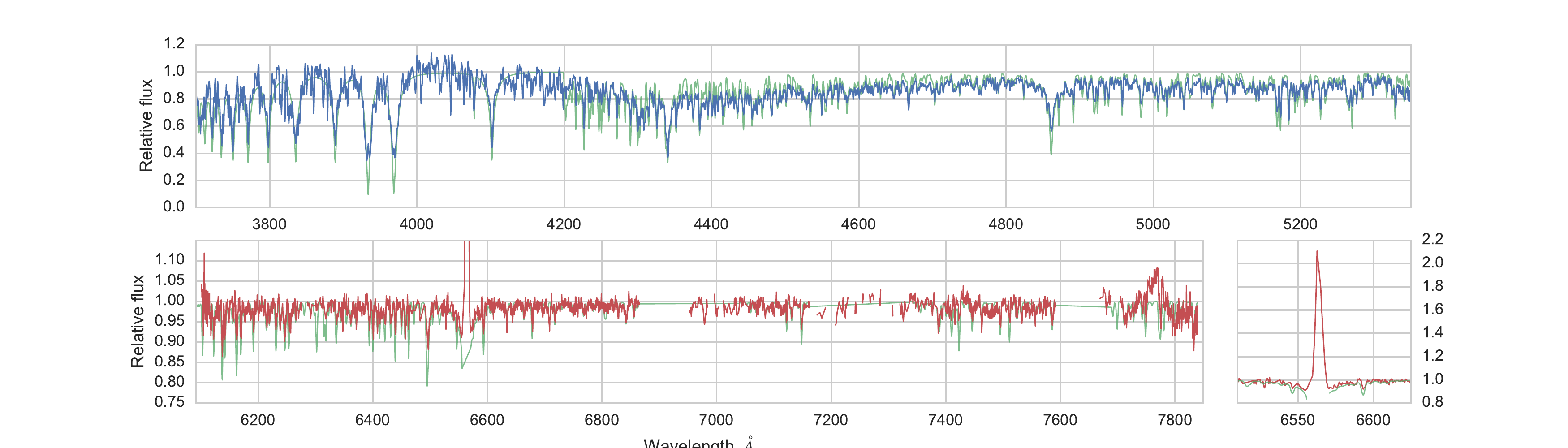}
\caption{Red and Blue spectra of PDS~110 taken with the WHT/ISIS. The best-fit synthetic spectrum is shown in green. H-alpha emission is shown in a separate plot in the lower right.}
\label{figure:whtspec}
\end{figure*}

\begin{figure}
\includegraphics[width=0.5\textwidth]{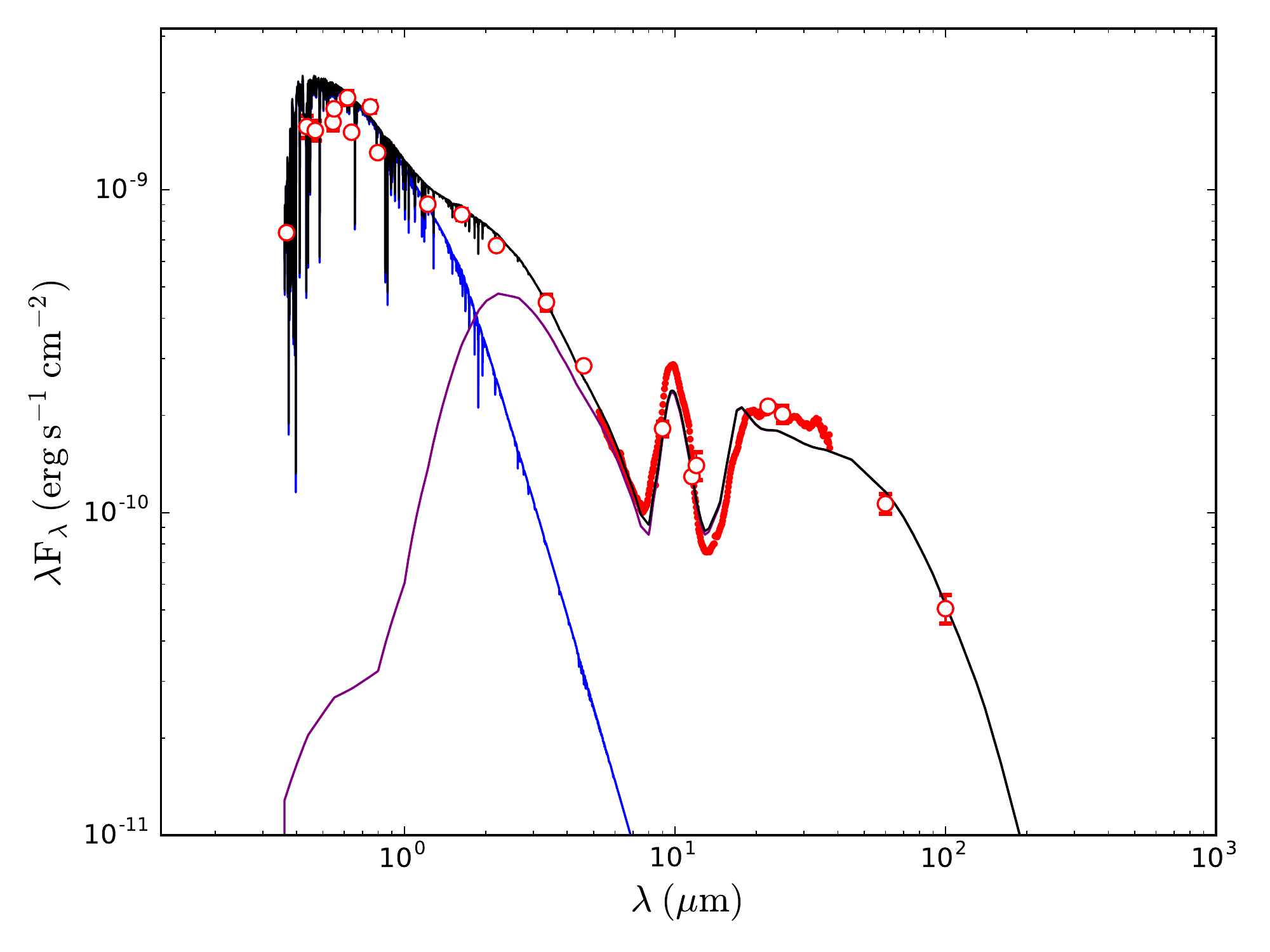}
\caption{Best-fitting model (black) to the SED of PDS~110. Photometry (red) are from Table 1 and {\it Spitzer} IRS \citep{Werner:2004, Houck:2004} low-resolution spectra are from the Cornell Atlas of {\it Spitzer} $\/$IRS Sources \citep[CASSIS,][]{Lebouteiller:2011}. The best-fitting model includes emission from a NextGen stellar photosphere \citep{Hauschildt:1999} (blue) and disc emission (purple).
}
\label{figure:SED}
\end{figure}

A low-resolution spectrum of PDS~110 was taken with the ISIS spectrograph in the R600B and R600R modes on the 4.2-m William Herschel Telescope at the ING, La Palma (shown in Figure \ref{figure:whtspec}).
The spectrum exhibits a strong H$\alpha$ emission line, 
moderate emission in the Ca H \& K line cores, and Li~I absorption at $\lambda=670.8$ \& 610.3~nm -- all consistent with previous measurements \citep{Torres:1995,Rojas:2008}. 
To characterise the spectra, a grid of 1200 synthetic spectra were generated with the Python package iSpec \citep{Blanco-Cuaresma:2014} using the MARCS model atmospheres \citep{Gustafsson:2008} and compared with the optical spectrum.
The best-fit models had $T_{\text{eff}}=6500\pm250$, log($g$)\, $\simeq$\, 3.8 and [Fe/H]$\,=\,-0.5\,\pm\,0.2$, in agreement with previous estimates of the stellar parameters \citep[e.g. 6653\,K, ][]{Gregorio-Hetem:2002}

\subsection{TRES spectra}
We have taken seven spectra of PDS~110 with the Tillinghast Reflector Echelle Spectrograph (TRES, \citet{Furesz2008,szentgyorgyi2007precision}) on the 1.5 m telescope at the Fred Lawrence Whipple Observatory (FLWO), Arizona. The TRES spectra have resolution R $\sim$ 44000 and were reduced, extracted, and analysed with the Spectral Parameter Classification (SPC) procedure of \citet{buchhave2012abundance}. 
We ran this without priors for each spectrum (with an average SNR of 53.5) and took a weighted average of the resulting stellar parameters. 
These give an effective temperature of T$_{\rm eff}=6360\pm110$K, a log~{\it g} of $3.89\pm0.17$ and [Fe/H]=$0.06 \pm 0.06$. 
Only metallicity shows a significant difference from previous estimates of stellar parameters. The higher precision of the TRES spectrum suggests this value is more precise, and we adopt it here.
The star is rapidly rotating with \hbox{$\upsilon \sin i_{\star}$} of $64.3 \pm 0.9$ km.s$^{-1}$

\section{Analysis}
\subsection{HR Diagram Position}
Previously published spectral types span F5-F7 \citep{Miroshnichenko:1999, Suarez:2006, Rojas:2008}.
Based on the \teff\, scale for pre-MS stars from \citet{Pecaut:2013}, a spectral type of F6 ($\pm$1 subtype) translates to a \teff\, estimate of 6250\,$\pm$\,140\,K.
Based on these estimates, we adopt a mean \teff\, of 6450\,$\pm$\,200\,K.

On the scale of \citet{Pecaut:2016}, this temperature translates to a $V$-band bolometric correction of BC$_V$ $\simeq -0.02$\,$\pm$\,0.02 mag. 
Fitting the $UBV$ photometry listed in Table 1 alone, the range of quoted spectral types translates to a reddening of E($B-V$) $\simeq$ 0.09 mag. 
Combining this estimate along with the two previous independent reddening estimates from \S2, we adopt a mean reddening estimate of E($B-V$) $\simeq$ 0.07\,$\pm$\,0.02 and $V$-band extinction of $A_V$ $\simeq$ 0.24\,$\pm$\,0.07 mag. 

Adopting the mean distance to the Ori OB1a from \citet{Hernandez:2005} as representative for PDS\,110, we can now calculate stellar parameters like absolute magnitude ($M_V$ = 2.54\,$\pm$\,0.11), apparent bolometric magnitude ($m_{bol}$ = 10.14\,$\pm$\,0.08), absolute bolometric magnitude ($M_{bol}$ = 2.52\,$\pm$\,0.11), luminosity (log(L/L$_{\odot}$) = 0.89\,$\pm$\,0.05 dex), and radius ($R$ = 2.23\,$\pm$\,0.18 $R_{\odot}$). Interpolating between the pre-MS isochrones from \citet{Siess:2000}, the stellar mass is 1.6\,M$^N_{\odot}$ and its age of $\sim$11 Myr, consistent with the rest of Ori OB1a. 

\begin{table}
\centering
\caption{Determined Stellar Parameters for PDS 110}
\label{tbl:Host_extract_props}
\begin{tabular}{|lcc|clc|}
\hline
$M_V$ & $2.54\pm0.11$ & ~& ~ & Teff & $6400\pm150$ K \\
$E(B-V)$ & $0.09$mag & ~& ~ & \logg & $3.8\pm0.3$\\
$A_V$ & $0.24\pm0.07$ & ~& ~ & $[\text{Fe/H}]$ & $0.06\pm0.06$\\
log($L/L_{\odot}$) &$0.89\pm0.05$ & ~& ~ & $R_s$ & $2.23\pm0.18 R_{\odot}$  \\
age & $\sim 11$ Myr & ~& ~ & $M_s$ & $\sim 1.6 M^N_{\odot}$ \\
\hbox{$\upsilon \sin i_{\star}$} & $64.3 \pm 0.9$  & ~ & ~ & ~ & ~ \\
\hline
\end{tabular}
\end{table}

\subsection{SED Disc model}
To model the SED of PDS~110, we used the self-consistent irradiated, accretion disc models of \cite{DAlessio:2006} to create a model grid using the stellar parameters of PDS~110 in Table 2.  We adopted a dust sublimation temperature of 1400 K to set the inner radius of the disc.  We included outer disc radii of 50 AU, 150 AU, and 300 AU, viscosity parameters (\textalpha) of 0.01, 0.001, and 0.0001 and dust settling parameters (\textepsilon; i.e. the dust-to-gas mass
ratio in the upper disc layers relative to the standard dust-to-gas mass
ratio) of 1.0, 0.5, 0.1, 0.01, 0.001, and 0.0001. 
The minimum grain size in the disc atmosphere was held fixed at 0.005 \textmu m while we varied the maximum grain size between 0.25, 1.0, and 2.0 \textmu m to reproduce the 10 \textmu m silicate emission feature.  The inclination angle was fixed at 60 degrees. 

Based on the $\chi$-squared values, the best fitting model has $a_{\text{max}}$=0.25$\mu =~m$, \textepsilon=0.5, \textalpha=0.01, and an outer radius of 300 AU. Uncertainties are beyond the scope of this analysis. This disc model has a mass of 0.006 $M^N_{\odot}$ using Equation 38 in \citet{DAlessio:1998}. 
While there are no millimeter data available to provide spatial constraints, a large outer radius of 300 AU is consistent with the significant MIR and FIR excess of this object given that the strength of the disc emission is related to the disc mass which in turn depends on radius \citep{DAlessio:1998}. We also note that \textepsilon=0.5 corresponds to a relatively flared disc.  Here we measure a disc height at 2 AU of 0.3 AU.

\subsection{Photometry}
Some out-of-eclipse variability is seen with peak-to-peak amplitudes on the order of $\sim$3\%. 
From the measured $v\sin(i)$ (64km.s$^{-1}$) we would naively estimate a P$_{\rm rot}$ of $\sim$1.7d.
However, searches with lomb-scargle periodograms \citep{PressFwls} on both the entire dataset and shorter segments do not detect any coherent period of variation attributable to rotation, with the signals dominated by day- and month- aliases from the ground-based surveys.
This suggests variations are stochastic or quasi-periodic, as has been seen for T-Tauri stars before \citep{rucinski2008photometric,Siwak:2011}.
The (space-based) IOMC light curves show candidate signals at 1.11d and 0.304d, with an amplitude of around 3\%. However, like the ground-based data, the time coverage is non-continuous.
The KELT light curves alone show a possible signal with P=67d.

Some dimmings, slightly shallower in depth and shorter in duration than the eclipses (only 3 to 4 points, or 9 to 12 days long) are also seen in ASAS data in 2006.
These are inconsistent with the proposed period (see section 4) seen and the lack of simultaneous data also means we are unable to rule out whether these are caused by systematics or from a genuine drop in stellar flux.

All observations thus far have also been monochromatic, with the ASAS, KELT and WASP data all focused on the V/R bands, and showing little differences in variability between one another in- or out-of-eclipse.


\subsection{Simple Eclipse Model}
\begin{figure}
\includegraphics[width=0.5\textwidth, clip, trim = 0 1in 0 6in]{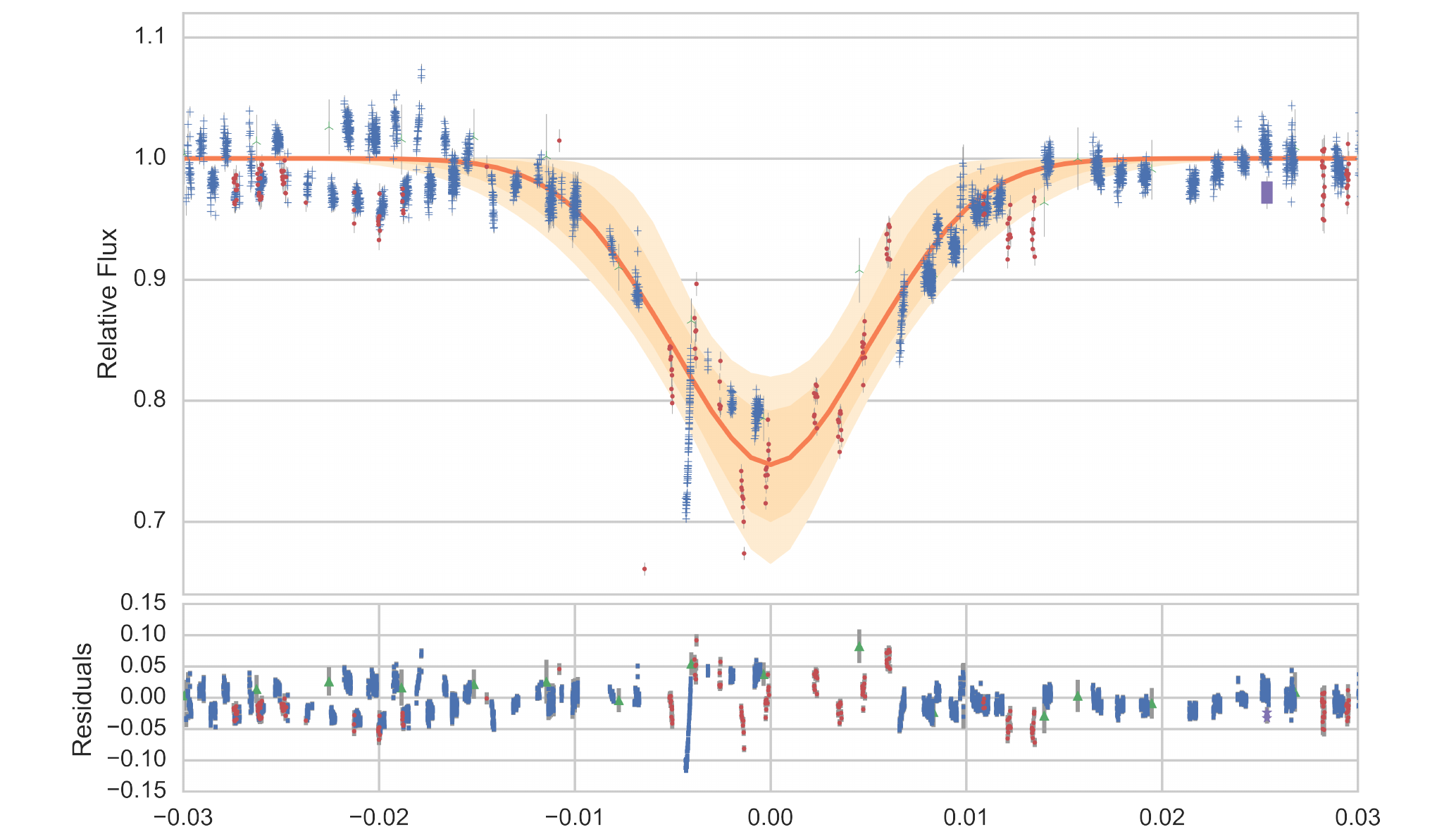}
\caption{
A best-fit Gaussian model compared to phase-folded data with 1-/2-sigma error regions in orange/yellow. Dark blue and red points show measurements from WASP and KELT respectively and have been phase-folded on the median period value.}
\label{figure:mcmcbestfit}
\end{figure}

The light curve clearly shows two significant dips in 2008 and 2011.
The implied period of \textasciitilde2.2 years means no observations were made during times of expected eclipses prior or after these two events.
The integer multiples of this separation (1.1 years, 0.73 years, etc) would produce observable eclipses in the current data, therefore have been ruled out.
%
We fit a simple Gaussian model to the phase-folded combined photometry to estimate the physical parameters of the eclipse.
An MCMC model was run to determine uncertainties on the best-fit with \texttt{emcee} \citep{ForemanMackey:2013} in \texttt{Python}.
The results of the model, and output posteriors, are shown in Figure \ref{figure:mcmcbestfit}.
We find the period to be $808\pm3$ days with an eclipse centred at HJD=$2454781\pm2$, depth of $26\pm6$\% and full-width half maximum of $7\pm2$ days.
The residuals show that the in-transit variability increases substantially compared to out-of-transit, indicative of finer structure in the eclipse light curve (see Figure \ref{figure:FullLC}).


\section{Interpretation and Discussion}
\label{sec:interp}





\subsection{Summary of Information}
We have detected two near-identical eclipses of the bright (V=10.4 mag), young (\textasciitilde 10Myr) star PDS-110 in the OB1a association with WASP, KELT and ASAS photometry.
The two events are separated by $808\pm2$ days and have nearly identical shapes, durations ($\sim 25$~days) and depths ($\sim 26\%$). Sharp in-eclipse gradients suggest fine structure in the eclipsing material.
The similarities of the eclipses strongly suggest that they are periodic.
Unfortunately, despite 25 seasons of data across 15 years and five surveys, all other predicted eclipses lie in observing gaps.

A study of the star and disc suggest that PDS-110 is a young Ge/Fe star surrounded by a thick protoplanetary disc which contributes to as much as \textasciitilde25\% of the total luminosity. 
Since we see significant optical emission and negligible extinction, we must be viewing the star at a significant inclination relative to the stellar disc.
Hence any eclipsing material must reside at a significant altitude above the disc midplane.

For any eclipse hypothesis we must take into account the shallow depth of this eclipse, its interesting substructure, and whether the material is optically thick or thin.
The most probable explanation is that the occulting object entirely crosses the star, but is optically thin.
In this case, the slow and Gaussian-like in- and egress gradients are the result of density gradients within a diffuse occulting dust cloud.
Sharp features during the eclipse can be explained as regions of sharply varying density within the cloud, such as gaps, clumps, thicker rings or ring gaps.
This would appear the most plausible scenario, although a mix of sharp optically thick regions and low-opacity dust regions, as has been proposed for the J1407b ring system, may also work. 
These scenarios can be disentangled with multiband photometry during eclipse (see Section 6).

There exist two potential mechanisms for the eclipses. First that circumstellar dust caused the eclipse; and second that the eclipse of a secondary body caused the eclipse. We explore these hypothesis in detail here.

\subsection{Circumstellar structure scenario}
Many processes within the large circumstellar dust disc could disturb dust above the midplane and into eclipse.
One possibility is from a spiral arm or a vortex. However, such scenarios are likely concentrated in the disc midplane, have significant azimuthal extent (of order radians), and move much more slowly than the material itself, hence not conducive to short, deep eclipses.

KH-15D-like dimmings, in which one member of a binary pair passes behind the circumstellar disc each orbit is another possibility. 
However, a binary on a ~2 year orbit would clear the entire inner disc region  - inconsistent with the disc model needed to explain the SED. 
If the total obscuration of a companion star leads to a $\sim$30\% dip during eclipse, it must be less than $-2.5*\log{0.3/0.7}$ or only $\sim$0.9 magnitudes fainter.
Hence such a bright companion would likely have been detected in either the CCF of the optical spectra or in the SED model.

Although the mechanism of eclipse remains unsolved, the deep and aperiodic dimmings or UX Ori stars (known as UXOrs), which are seen around many Herbig Ae/Be stars \citep{Bertout:2000} are similar to the eclispes seen in PDS-110.
Some eclipses resemble a single PDS-110 eclipse in depth or duration \citep[e.g.][]{Caballero:2010}.
However, these dimming events tend to be deeper (often several magnitudes), longer-duration (weeks to years) and are aperiodic.
Lightcurves of those UXOrs found also tend to exhibit many events, usually with differing depths and durations. 

The proposed mechanisms for UXOr-like dimming events include hydrodynamical fluctuations at the inner edge of self-shadowed circumstellar discs \citep{Dullemond:2003}, occultations of dust clumps in their circumstellar disc \citep[][etc.]{Grinin:1988,Voshchinnikov:1989,Grady:2000}, and the eclipsing debris of planetesimal collisions in young asteroid belts \citep{kennedy:2017}.
As an F-type star there is no guarantee that the self-shadowing proposed as a cause of UXOr behaviour is present for PDS-110.
Our tentative SED fit also suggests an unsettled ($\epsilon$=0.5) and moderately turbulent ($\alpha$=0.01) disc - atypical for UXOrs \citep{Dullemond:2004}.
Regardless, the inferred period for the events, and their rarity, suggests the occulting material lies well beyond the disc's inner edge at the sublimation radius.
The lack of other significant variability suggests that the occulting material lies well above the "main" disc, and that the disc structure may be relatively unimportant for determining the nature of the eclipses. 

While this style of eclipse does not fit what is observed for PDS-110, it is possible that we could be observing a new UXOr-like eclipse behaviour.

Regardless of the formation mechanism, any diffuse clumps would be subject to shear. 
The angular shear rate is R$d \Omega / dR = - 3 \Omega / 2$, so across a clump of radius $R_{\rm cl}$ the shear velocity is $v_{\rm sh} = 6 \pi R_{\rm cl} / P$ (where $P$ is the orbital period). 
That is, a clump of any size will be sheared out by a factor of $6\pi$ after one orbit, and the radial and vertical optical depth will be roughly $6\pi$ times lower. 
Any disc structure will shear out rapidly, and on successive orbits will have a very different azimuthal extent. 
Thus, the similar shapes of the eclipses mean that if they were caused by the same clump, an additional means to maintain the concentration of this clump is needed. 

\subsection{Circumsecondary structure scenario}
We have therefore established that the eclipsing object is highly likely to be periodic, and unlikely to be formed of streams or clumps of dust.
The simplest way of concentrating material is with the gravitational attraction of a massive body.
This is the established interpretation for many long-duration eclipses in young systems, with orbits from 48d \citep{Herbst:1994} to \textasciitilde70 years \citep{Rodriguez:2016A}.
We explore here the likely characteristics of such a body by considering its Hill Sphere.

\subsubsection{Hill Sphere Considerations}
With an orbital period of P=808 days, and a total mass of 1.6 solar masses, we derive a circular Keplerian velocity of 27\kms.
By assuming an eclipse is caused by an optically-thick knife-edge moving across the star, the gradient of the steepest slope can be used to give a minimum velocity of the eclipsing object.
For the sharp flux increase seen at 2008.85 ($\sim$20\% in 6 hours) in WASP data (Figure 1), this gives $v_{\rm min} = 13$\kms. 
This is therefore consistent with the implied orbital motion of 27\kms as an optically thin or angled structure could produce faster velocities for a given slope.
From the Keplerian orbital speed and eclipse duration ($\sim$21 days), we can derive the diameter of the eclipsing object to be $\sim$0.3\,au, or about 50 million km.
A lower limit on the mass of the secondary companion can be derived assuming that the cloud is within the Hill sphere of the secondary.

%
%
%

%

The Hill radius can be approximated as: $a_H \approx a \left( 1-e \right) \left( m / 3M_s \right) ^{1/3}$ where $a$ is the orbital semi major axis, $e$ is the orbital eccentricity, $m$ is the mass of the secondary and $M_s$ is the primary (stellar) mass.
If the duration of the eclipse is $t_{ecl}$ days, then the diameter of the disc $ d_{disc} = v_{circ} t_{ecl}$ and the circular velocity of the companion $v_{circ}=2\pi a/P$.
Combining these expressions with Kepler's third law, the mass of the secondary companion is:
$ m = 2M_s \left( \pi t_{ecl} / \xi P \right) ^3 $
 where $P$ is the orbital period of the secondary companion and $0 \leqslant \xi \leqslant 1.0$ is the fraction of the Hill sphere that the disc fills. $\xi=0.3$ is typical for a prograde rotating disc \citep{Nesvorny:2003}.

Assuming M$_s$=1.6M$^N_\odot$ and $t_{ecl}=21d$ and $P=808$ days gives: $m = 1.8 M_{Jup}\left( 1 / \xi \right)^3 $
Using the prograde stability criterion of $\xi\,=\,0.3$ \citep{Quillen:1998}, the mass is 
68\,M$_{Jup}$ and for $\xi$\,=\,0.6, this becomes
8.5\,$M_{Jup}$.
Increasing the eclipse duration (for example, by including the shallow dips seen 15-20 days before and after) will substantially increase this mass limit (to $>20M_{Jup}$ in the case of a 40~day eclipse).


Such a body would likely also perturb a gap in the circumstellar disc at 2.2AU.
We recomputed the SED model with a narrow gap at this radius and found it to be consistent with the data, with negligible difference to a gapless model.

\subsubsection{Inclination Considerations}
The two eclipses have similar duration of \textasciitilde25 days and so we assume that the cloud has a constant size. 
We hypothesize that the eclipse is caused by the passage of a large optically thin cloud that contains an unseen secondary companion which holds the cloud together in dynamic equilibrium, and that the companion and cloud orbit around the primary.

In the cases of KH-15D, $\epsilon$ Aur and EE Cep the secondaries are stars, whereas for J1407 the massive body at the centre of the disc appears to be of planetary or brown dwarf mass.
In order to cause the eclipse, either:

\textbf{(1)} The secondary body is large and on an orbit with low mutual inclination to the disc, but with highly inclined (Uranus-like) circumsecondary material which protrudes above the circumprimary disc and passes our line-of-sight of the primary star.
If, as our lack of reddening suggests, we are viewing PDS-110 at an angle moderately inclined from edge-on ($\sim30^{\circ}$), the eclipsing secondary disc must be greater than $\sim$1AU in radius, hence stellar in mass. This, it would likely be detected as anomalous flux in the optical spectra and SED fit.

\textbf{(2)} The secondary body is small but has significant orbital inclination with respect to the disc. Such an orbital scenario could occur due to scattering.
This is our favoured scenario, and would be invisible except during eclipse.
A figure representing this scenario PDS~110 system is shown in Figure \ref{figure:pds}.


\begin{figure}
\includegraphics[width=0.5\textwidth]{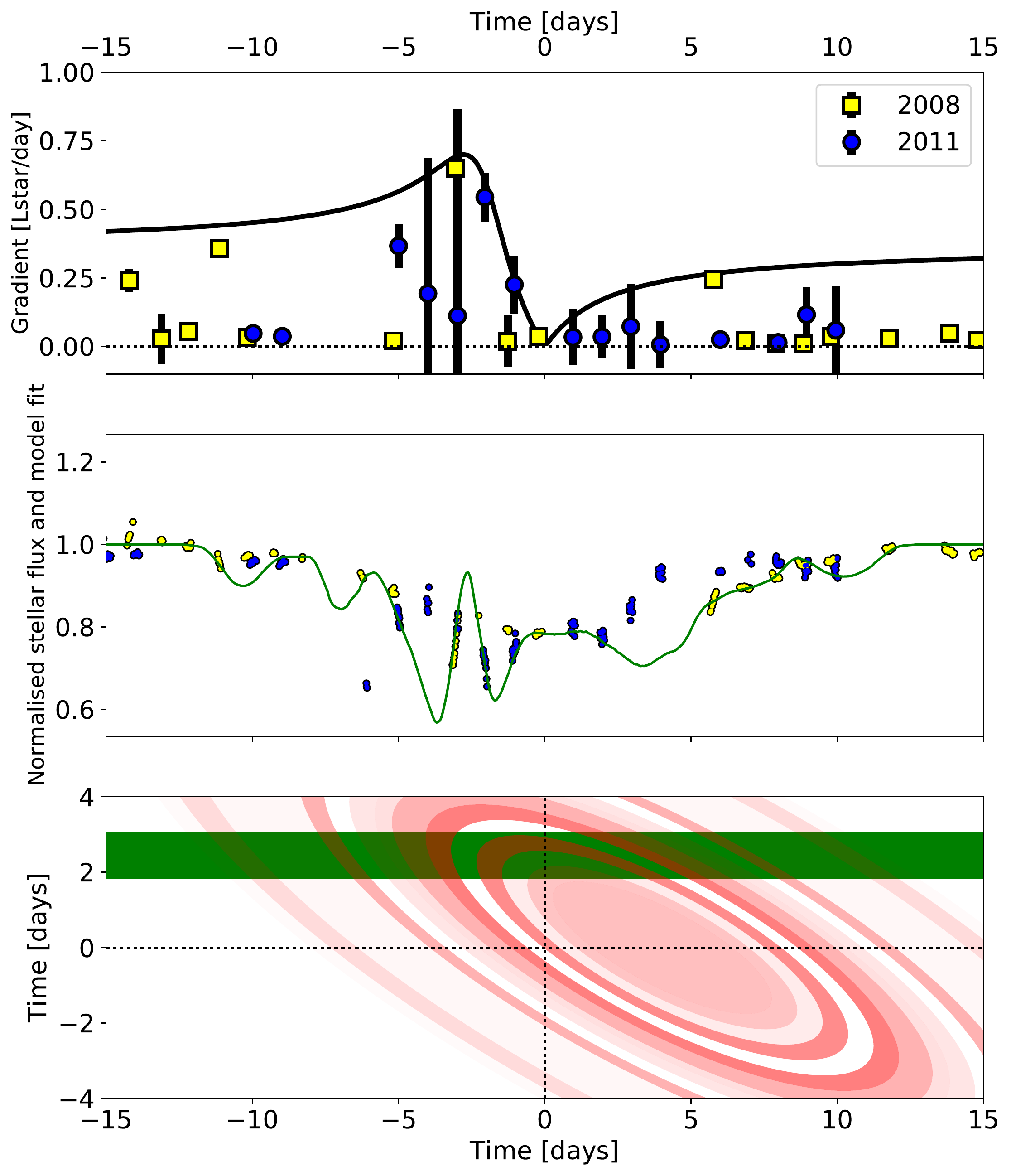}
\caption{
A circumsecondary ring model for PDS 110b. \textbf{Upper panel:} The light curve gradients seen in 2008 and 2011 photometry (yellow and blue) are shown, along with an upper bound fit to the gradients (black) from which the orientation of the system is derived. \textbf{Central panel:} The photometry in 2008 and 2011 together with the one plausible ring transmission model (green line).  \textbf{Lower panel}: A schematic of the model ring system (red nested ellipses) crossing the stellar disc (green).
}
\label{figure:ringmodel}
\end{figure}



\subsubsection{Circumplanetary Ring Model}

The WASP eclipse shows substructure over individual nights in the form of steep gradients similar to those seen towards J1407 (Mamajek et al. 2014).
While the interpretation is uncertain, we briefly consider the implications of a circumplanetary ring model using the framework of \citet{Kenworthy:2015}.
The rapid changes seen in eclipse, reminiscent of J1407 \citep{Mamajek:2012}, could be interpreted as the passage of a Hill sphere filling ring system around a secondary companion, passing across the line of sight of the star. 
To explore whether such a mechanism could explain the PDS~110 eclipse, we applied the exoring fitting method of \citet{Kenworthy:2015} to the WASP eclipse light curve.

\begin{figure*}
\centering
\includegraphics[width=0.7\linewidth]{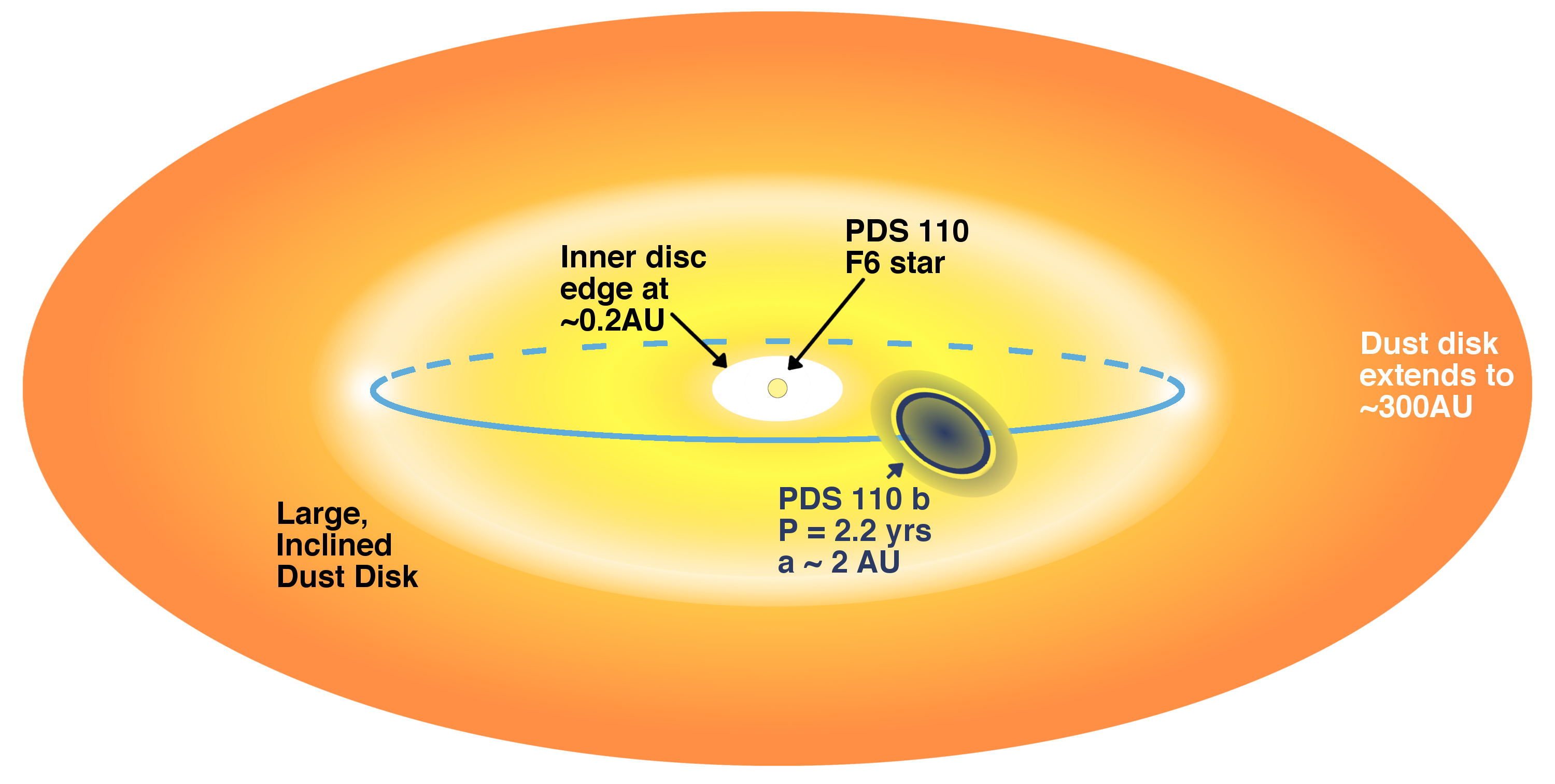}
\caption{A sketch of the PDS~110 system. The primary star is surrounded by a warm disc of dust inclined away from edge-on. Orbiting around the primary star is a secondary companion with an extended disc which eclipses the primary every 808 days.
}
\label{figure:pds}
\end{figure*}
We consider the light curve slopes in the WASP and KELT light curves separately.
We set the midpoint of the WASP eclipse as 2454780.7 days, as determined by the Gaussian fit carried out in Section 4.2, and we take the centre of the KELT eclipse to be at 2455590.4 days, determined by visual inspection of the two light curves and adjusting them so that the photometry of the different epochs gives the most consistent match in both photometry and in the matching of the light curve gradients.
The measurements of light curve gradient as a function of time from the centre of the respective WASP and KELT eclipses are shown in Figure \ref{figure:ringmodel}. 
The figure shows that there are seven light curve gradients above 0.1 Lstar/day during the ingress of PDS~110b, compared to only one during egress.
WASP detects 5 slopes and KELT detects three significant slopes.
From this we conclude that many more steep gradients are seen during ingress in both eclipse events, and in the context of the ring models, this implies an eclipsing object with small spatial scale structure similar to that seen in J1407b.
These gradients are used to determine the orientation of the ring system following the method of \cite{Kenworthy:2015} (Section 3.1).
By fitting the measured gradients g(t) to the model gradients G(t) we achieve the fit shown in Figure \ref{figure:ringmodel} as the solid black curve. 
All gradients must lie on or below this curve for there to be a consistent ring model. 
The determined disc parameters have an Impact parameter of 2.45 d, a ring center offset of 4.02 d (both in velocity space), an apparent disk inclination of 74$^\circ$, and total obliquity of the disk plane to the orbital plane of 26$^\circ$.

%

We then model the ring radii and transmissions as the convolution of the stellar disc (R=$2.23R_{\odot}$) with the ring parameters. 
Minimization of ring transmissions produced the ring model as seen in Figure \ref{figure:ringmodel}

The incomplete coverage of both eclipse events leads to several plausible ring solutions, of which we show just one in Figure 5. The ring model fits both epochs well in several places, and shows deviations in others. 
From the plot of light curve gradients, where we can see several high gradients on the ingress of the transit in both epochs, we conclude that a tilted disk containing azimuthal structure at high spatial frequencies is a reasonable fit to the data.
There isn't a unique solution using azimuthally symmetric structures, which may be due to several causes: (i) We are seeing at different clocking angles in successive transits (eg from a spiral pattern), (ii) the intrinsic stochastic variability of the parent star is affecting the derived photometry and light curve gradients, (iii) precession of the disk between successive transits, (iv) the eclipses are instead aperiodic dimmings caused by unexplained dust disc processes.
A comprehensive photometric monitoring campaign during future eclipses will help resolve these ambiguities in the interpretation of this object.


\section{Future Observations}
While we favor the presence of dust structure around a periodic secondary companion as the cause of the eclipses, additional data is needed to test it.  In particular, the next eclipse will take place on HJD=$2458015.5 \pm10$ (Sept 9-30 2017).
Unfortunately, it will only be observable for a few hours each night from the ground, and space based observations may be needed for better temporal coverage of the event.
The presence or absence of an eclipse will immediately settle the question of periodicity.

High cadence and low noise light curves during the eclipse will better constrain the presence of any smaller scale structures in the eclipsing material, potentially confirming the hypothesis that it is a disc of material with gaps and other structures orbiting a low mass secondary. 
Color information during the eclipse can determine if the obscuration is due to small dust grains or larger bodies that produce more achromatic absorption. 
The continuing out-of-eclipse monitoring by photometric surveys may detect other eclipsing structures and further characterize any other variability.  
A secondary should produce radial velocity variations in the primary of $ \sim 200$~m/s (for a $10 M_{Jup}$ companion) that may be measurable.
The difficulty is that the fast rotation and variability of the primary will limit the precision of radial velocity measurements.
While the scales corresponding to the orbit of the potential secondary ($\sim 2$~AU) cannot be resolved in sub-millimeter observations, they can characterize the disc on larger scales (10s of AU) and search for distortions or gaps in the outer disc that might indicate the presence of other massive bodies in the system.

\section{Conclusions}
We have detected two near-identical eclipses of the bright (V=10.4 mag), young (\textasciitilde 10Myr) star PDS-110 in the OB1a association with WASP, KELT and ASAS photometry.
Further ASAS-SN and IOMC photometry of the star have increased the photometric coverage of this star to 25 seasons of data across 15 years.
We interpret these eclipses to be caused by the same optically-thin clump of material on a $808\pm2$ day orbit around the star.

Despite a large circumstellar disc around PDS-110, such a scenario cannot be caused by lose clumps of dust above the disc plane, as shearing forces would not maintain the eclipse depth and duration across 2.2 years.
Therefore, we interpret the eclipse structure to be gravitationally bound around a companion body, which must have mass $>1.8M_{Jup}$.

Such a body must be significantly inclined relative to the circumstellar disc to eclipse the star.
The body may be surrounded by rings, as has been hypothesised for J1407, with the sharp photometric gradients seen at $t_0 \pm5$ days being the result of the transit of a ring gap.
This hypothesis can be tested, and the orbiting body studied in much greater detail, in September 2017 when we predict the next eclipse to take place.




\section{Acknowledgments}

The authors thank the Lorentz Center at Universiteit Leiden
for supporting the "Rocks, Rubble and Rings: Understanding Deep and Irregular Transits" workshop in 2016.
HPO was funded by a University of Warwick Chancellor's Scholarship.
Work performed by J.E.R. was supported by the Harvard Future Faculty Leaders Postdoctoral fellowship.
GMK is supported by the Royal Society as a Royal Society University Research Fellow.
Early work on KELT-North was supported by NASA Grant NNG04GO70G. J.A.P. and K.G.S. acknowledge support from the Vanderbilt Office of the Provost through the Vanderbilt Initiative in Data-intensive Astrophysics. This work has made use of NASA's Astrophysics Data System and the SIMBAD database operated at CDS, Strasbourg, France.
This paper is under review for unlimited release (URS265682).
D.J.A. acknowledges funding from the European Union Seventh Framework programme (FP7/2007- 2013) under grant agreement No. 313014 (ETAEARTH). 
Work by B.S.G. was partially supported by NSF CAREER Grant AST-1056524. Work by K.G.S. was supported by NSF PAARE grant AST-1358862. 
E.E.M. acknowledges support from the NASA NExSS program. 
Part of this research was carried out at the Jet Propulsion Laboratory, California Institute of 
Technology, under a contract with the National Aeronautics and Space Administration. 
TW-SH is supported by the DOE Computational Science Graduate Fellowship, grant number DE-FG02-97ER25308.
C.S.K. and K.Z.S.  are supported by NSF grants AST-1515876 and AST-1515927.
%





\bibliographystyle{mnras}

\bibliography{PDS110}

\end{document}